\documentclass[aps,prd,showpacs,amsmath,amssymb,reprint,superscriptaddress]{revtex4-1}
\usepackage{hyperref}
\usepackage{graphicx}
\usepackage{color}
\usepackage{subfigure}

\def\be{\begin{equation}}
\def\ee{\end{equation}}
\def\bea{\begin{eqnarray}}
\def\eea{\end{eqnarray}}

\def\lsim{\, \rlap{$<$}{\lower 1.1ex\hbox{$\sim$}}\,}

\begin{document}

\title{Greybody factors for non-minimally coupled scalar fields in
Schwarzschild-de~Sitter spacetime}

\author{Lu\'is C. B. Crispino}
 \email{crispino@ufpa.br}
  \affiliation{Faculdade de F\'isica, Universidade Federal do Par\'a,
66075-110, Bel\'em, PA, Brazil}

\author{Atsushi Higuchi}
 \email{atsushi.higuchi@york.ac.uk}
  \affiliation{Department of Mathematics, University of York,
Heslington,
York YO10 5DD, United Kingdom}

\author{Ednilton S. Oliveira}
 \email{ednilton@ufpa.br}
 \affiliation{Faculdade de F\'isica, Universidade Federal do Par\'a,
66075-110, Bel\'em, PA, Brazil}

\author{Jorge V. Rocha}
 \email{jorge.v.rocha@ist.utl.pt}
 \affiliation{CENTRA, Dept. de F\'{\i}sica, Instituto Superior T\'ecnico, Technical University of Lisbon,
 Avenida Rovisco Pais 1, 1049 Lisboa, Portugal}

\date{\today}

\begin{abstract}
We compute the greybody factors for
non-minimally coupled scalar fields in four-dimensional Schwarzschild-de~Sitter spacetime.
In particular, we demonstrate that the zero-angular-momentum greybody factor generically tends to zero in the zero-frequency limit like frequency squared if there is non-vanishing coupling to the scalar curvature.
This is in contrast with the minimally coupled case, where the greybody factor is known to tend to a finite constant.
We also study the Hawking radiation for non-minimally coupled massless scalar fields in Schwarzschild-de~Sitter spacetime, formulate a sensible notion of a generalized absorption cross section and investigate its properties.
\end{abstract}


\maketitle

\section{Introduction}

The study of phenomena in de Sitter spacetime is a subject of importance beyond pure academic interest, given that our cosmological neighborhood is presently undergoing accelerated expansion~\cite{Supernovae}.
Moreover, de Sitter (dS) spacetime is a very good approximation to the exponentially expanding phase postulated by the inflationary paradigm~\cite{Guth:1980zm}.
Finally, possible connections with conformal field theories provided by the dS/CFT correspondence~\cite{Strominger:2001pn} add value to the study of asymptotically dS spacetimes.
On the other hand, black holes are among the most relevant objects in any gravitational theory, besides there being clear indications of their presence at the center of galaxies~\cite{Begelman:2003xt} and the possibility of black hole (BH) formation in particle colliders~\cite{MiniBHs}.
These observations motivate the study of black holes in asymptotically dS spacetimes.

Most of the literature investigating the scattering and absorption properties of waves in BH spacetimes focus on the asymptotically flat case.
Recently, such problems have been studied extensively~\cite{Doran_etal, Dolan_etal, Dolan1,cohm,cdo2,cdo1,cho}.
In asymptotically flat spacetimes it is common to express the outcome of scattering a wave off a black hole in terms of an absorption cross section. The absorption cross section is directly connected to the \emph{greybod factors},
i.e., the probability for a given wave coming in from infinity to be absorbed by the black hole~\cite{Myung:2003cn, HNS_atmp14_727}.
This has been shown to be equal to the transmission probability for an outgoing wave `emitted' from the black hole event horizon to reach the asymptotic region (see e.g.\ Ref.~\cite{HNS_atmp14_727}).
It is the non-triviality of greybody factors that cause the semi-classical spectrum of emission of black holes to depart from that of a pure black body.

For asymptotically flat BH spacetimes there are indications that the greybody factor $\gamma_l(\omega)$ for waves of arbitrary spin $s$ and angular quantum number $l$ in any number of dimensions $d$ vanishes in the zero-frequency limit.
This general statement is confirmed by all cases studied so far, and has actually been proven for massless minimally coupled scalar fields in stationary black hole backgrounds~\cite{prl78_417,Atsushi} in general spacetime dimensions.
In $d=4$ this phenomenon also occurs for massless spin-1/2 fermions, gauge bosons and gravitons~\cite{prd13_198} and persists even in the presence of non-minimal coupling of the scalar field with the curvature scalar~\cite{Chen:2010ru}.

However, the $l=0$ greybody factor tends to a positive constant in the infrared limit for a minimally coupled massless scalar field in Schwarzschild-de~Sitter spacetime (SdS).
The expression obtained in the zero-frequency limit in four dimensions was first reported in Ref.~\cite{BCKL}, and the result for arbitrary dimensions is~\cite{KGB}
\be
\gamma_0(\omega=0) = \frac{4(r_C r_H)^{d-2}}{\left(r_C^{d-2}+r_H^{d-2}\right)^2}\,,
\ee
where $r_C$ and $r_H$ stand for the radial location of the cosmological horizon and the black hole event horizon, respectively.
Several other authors have confirmed this result, e.g.~\cite{Wu:2008rb, Liu:2010ar, HNS_atmp14_727}.
This means that at low energies the cosmological constant has an important effect on the greybody factor of massless minimally coupled scalar particles.
The explanation put forward in~\cite{KGB} for this phenomenon is that zero-energy particles are fully delocalized, and have therefore a finite probability to traverse the distance between the two horizons~\cite{footnote}.
This argument then suggests that the infrared enhancement of transmitted flux in SdS is not present when fields are massive, since the dispersion relation gets modified by the addition of the mass.
In this paper we will indeed find that this phenomenon is specific to the massless case.

To the best of our knowledge, the greybody factors for massless scalar fields propagating in the SdS geometry have been computed only for minimal coupling.
In this paper we consider a massless scalar field with nonzero coupling to the scalar curvature, $\xi R\phi^2$,
propagating in 4-dimensional SdS spacetime and compute the corresponding greybody factors.
Note that this coupling can also be interpreted as a mass term since the scalar curvature of the background is constant
in SdS spacetime.
We show that, if $\xi\neq 0$, the greybody factor for the $l=0$ mode tends to zero in general like $\omega^2$, where $\omega$ is the frequency of the wave.
We also show that the rate of Hawking radiation tends to zero as $\omega\to 0$ if $\xi\neq 0$, unlike the case with $\xi=0$, in which this rate remains finite in the low-frequency limit.

The definition of an absorption cross section in dS spacetimes
is an issue that has generated some controversy~\cite{HNS_atmp14_727}.
We elucidate that this concept, as it is defined in Ref.~\cite{KGB}, is generally not meaningful in dS spacetimes.
However, for small black holes ($r_C \gg r_H$) it is possible to define a {\em generalized} absorption cross section, albeit only approximately.

The remainder of this paper is organized as follows.
We briefly describe Schwarzschild-de~Sitter spacetime in Sec.~\ref{sds} and present the framework for finding the behavior of a massless scalar field in this spacetime for general coupling $\xi$.
In Sec.~\ref{gbf_sec} we explain how the greybody factor is computed and find its low-frequency limit for $\xi\neq 0$ to second order in $\xi$.
We also derive a complementary low-frequency approximation, valid only for small SdS black holes but for arbitrary $\xi$.
To conclude this section we present numerical results for the greybody factors with several values of $\xi$.
Hawking radiation is analyzed in Sec.~\ref{ee} and some properties of the generalized absorption cross section are studied in Sec.~\ref{acs}.
We summarize our results in Sec.~\ref{conclusion}.
In Appendix~\ref{swp} we briefly discuss the scattering problem in a negative square-well potential, which may be helpful in understanding the special nature of minimal coupling $\xi=0$.

We use natural units such that $c = G = \hbar = k_B = 1$ and metric signature $(-,+,+,+)$.

\section{Schwarzschild-de~Sitter spacetime}
\label{sds}

The 4-dimensional Schwarzschild-de~Sitter spacetime has line element given by
\begin{equation}
 ds^2 = -f(r)dt^2 + f(r)^{-1} dr^2 + r^2 (d\theta^2 + \sin^2 \theta d\phi^2),
\label{ds}
\end{equation}
where
\begin{equation}
 f(r) = 1 - \frac{2M}{r} - \frac{\Lambda}{3} r^2 ,
\label{f}
\end{equation}
with $M$ being the black hole mass and $\Lambda >0$.
The metric $g_{\mu\nu}$ defined by $ds^2=g_{\mu\nu}dx^\mu dx^\nu$ is a solution of the vacuum Einstein field equations with positive cosmological constant $\Lambda$,
\begin{equation}
R_{\mu \nu} - \frac{1}{2} g_{\mu \nu} R + \Lambda g_{\mu \nu} = 0 \,,
\label{EFE}
\end{equation}
where $R_{\mu \nu}$ and $R$ denote the Ricci tensor and scalar curvature, respectively.

The roots of $f $ occur at $r = r_H, r = r_C$ and $r = -r_H - r_C$, where $r_H$ and $r_C$ are the event and cosmological horizons of the SdS spacetime, respectively (with $r_H<r_C$).
It is easy to see that the metric~\eqref{ds} depends only on a single dimensionless parameter, which is commonly taken to be $\lambda\equiv \Lambda M^2/3$, up to an overall constant scale factor. The spacetime features two horizons only for $0<\lambda<1/27$~\cite{Gibbons:1977mu, Stuchlik:1999qk} (see~\cite{Molina:2003ff} for the higher-dimensional case).
For $\lambda=1/27$ the two horizons merge and one obtains the extreme SdS geometry, whereas if $\lambda>1/27$ there are no horizons.
In the present study we shall restrict ourselves to the case $0<\lambda<1/27$.

The temperatures associated with the black hole event horizon and the cosmological horizon of the four-dimensional SdS solution are given by~\cite{Gibbons:1977mu, KGB}
\bea
T_H &=& \frac{1}{4\pi r_H} \left( 1 - \Lambda r_H^2 \right)\,,
\label{th}\\
T_C &=& \frac{1}{4\pi r_C} \left( \Lambda r_C^2 -1\right)\,.
\label{tc}
\eea
We note that the temperature of the black hole is always larger than the temperature of the universe ($T_H>T_C$), since $r_H<r_C$, implying a net energy flow from the black hole horizon towards the cosmological horizon~\cite{KGB}.

\subsection{Non-minimally coupled massless scalar field}
\label{sf}

We consider a massless Klein-Gordon field $\Phi(x^{\mu})$ in the background geometry~\eqref{ds}, coupled to the gravitational field, described by the action
\begin{equation}
S = -\frac{1}{2} \int
d^4 x \, \sqrt{-g} \, \left[
\left( \nabla_{\mu} \Phi \right) \,
\left( \nabla^{\mu} \Phi \right)
+ \xi R \Phi ^2
\right] \,,
\label{Action}
\end{equation}
where $g$ is the determinant of the spacetime metric and $\xi$ is the coupling between the scalar and the gravitational fields.
Some particular values of $\xi$ are of specific interest:
$\xi=0$ is the so-called minimal coupling, whereas $\xi=1/6$ is the so-called conformal coupling for which the scalar field theory becomes conformally invariant~\cite{BD}.

The massless scalar field satisfies the Klein-Gordon equation,
\begin{equation}
\left[  \nabla_{\mu}  \nabla^{\mu} - \xi R  \right] \,
 \Phi (t,r,\theta,\phi)
 = 0.
\label{KG}
\end{equation}

Taking advantage of the spherical symmetry of the problem and of the existence of the Killing vector field $\partial_t$,
we write the solutions to Eq.~(\ref{KG}) in the form
\begin{equation}
 \Phi (t,r,\theta,\varphi) = \frac{\psi_{\omega l}(r)}{r} Y_{lm} (\theta,\varphi)
e^{-i\omega t}, \,\,\,\,\,\,\,\, \omega >0 \, ,
\label{Phi}
\end{equation}
where $Y_{lm} (\theta,\varphi)$  are the scalar spherical harmonics.
The radial part of the waves obeys the following equation:
\begin{equation}
-f\frac{d}{dr} \left[f\frac{d}{dr} \psi_{\omega l} (r) \right] +
V(r)  \psi_{\omega l} =
\omega^2 \psi_{\omega l},
\label{radial_eq}
\end{equation}
where the effective potential $V(r)$ is given by:
\begin{equation}
 V(r) = f(r) \left[\frac{f'(r)}{r} +
\frac{l(l+1)}{r^2}
+ 4 \xi \Lambda
\right],
\label{potential}
\end{equation}
with the ``prime'' standing for the derivative with respect to $r$.

\begin{figure}
\centering
\includegraphics[width=8.6cm]{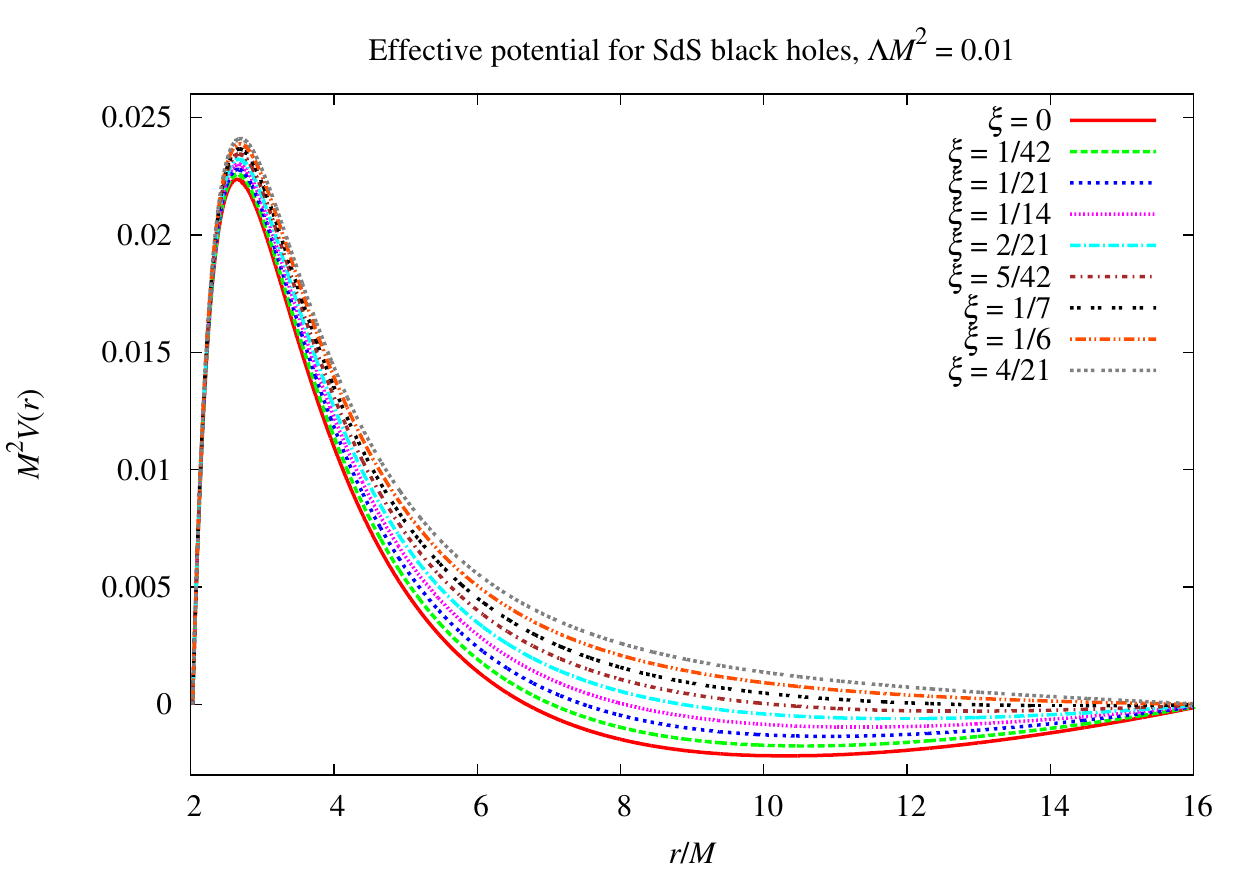}
\caption{Effective potential plotted for $\Lambda M^2 = 0.01$ and $l = 0$, for
different choices of the coupling parameter $\xi$.}
\label{veff}
\end{figure}

In FIG.~\ref{veff} we plot the effective potential~(\ref{potential}) for $\Lambda M^2 = 0.01$, $l = 0$
and for several values of $\xi$.
This effective potential vanishes at the SdS black hole horizon $r_H$ and at the cosmological horizon $r_C$.
It is apparent that the effective potential presents different behaviors for different values of the coupling parameter $\xi$.

\section{Greybody factor}
\label{gbf_sec}

\subsection{Asymptotic solution and greybody factor}

In order to determine the greybody factors, we need to know the asymptotic behavior
of the radial function $\psi_{\omega l}(r)$.

Since $V(r)$ goes to zero at the black hole event horizon, near $r=r_H$ we can write
\begin{equation}
 \psi_{\omega l} (r) \approx A_{\omega l}^{\text{tr}} e^{-i\omega r_*},
\label{asy_hor}
\end{equation}
with $r_*$ being the tortoise coordinate defined by $dr_*/dr\equiv f^{-1}$.
We imposed purely ingoing boundary conditions at the black hole horizon, thus eliminating a term proportional to $e^{+i\omega r_*}$ from the general solution.
We may write $r_*$ as:
\begin{equation}
r_* = \sum\limits_{n=1}^{3} \frac{1}{f'(r_n)}\log |1 - r_n/r|\,,
\label{tortoise}
\end{equation}
where $r_n$ denotes the roots of $f$.

As we approach the cosmological horizon, again $V(r)$ goes to zero and therefore
\begin{equation}
 \psi_{\omega l} \approx A_{\omega l}^{\text{in}} e^{-i\omega r_*} + A_{\omega l}^{\text{out}} e^{+i\omega r_*},
 \label{psi_rc}
\end{equation}
where $A_{\omega l}^{\text{in}}$ represents the amplitude of the incoming wave,
while $A_{\omega l}^{\text{out}}$ stands for the outgoing wave.
Hence, the greybody factors can be expressed as
\begin{equation}
\gamma_l(\omega) = \left|\frac{A_{\omega l}^{\text{tr}}}{A_{\omega l}^{\text{in}}} \right|^2\,.
\label{gbf}
\end{equation}
Since $|A_{\omega l}^{\text{in}}|^2 = |A_{\omega l}^{\text{tr}}|^2 + |A_{\omega l}^{\text{out}}|^2$ (from flux conservation), the greybody factors can also be written as
\begin{equation}
\gamma_l(\omega) = 1 - \left|\frac{A_{\omega l}^{\text{out}}}{A_{\omega l}^{\text{in}}} \right|^2.
\label{gbf2}
\end{equation}
We will also refer to the greybody factor as transmission coefficient.

\subsection{Low-frequency approximation with small $\xi$}


It was shown in Ref.~\cite{BCKL} that for $\xi=0$ the transmission coefficient for $l=0$ is finite and given by
\begin{equation}
 \lim_{\omega\to 0}\gamma_0(\omega) =
 \frac{4 r_H^2 r_C^2 }{\left( r_H^2 + r_C^2 \right)^2}.
\label{gbfmc0}
\end{equation}
However, as we stated in the introduction, if $\xi\neq 0$ the transmission coefficient $\gamma_0(\omega)$ generally behaves like $\omega^2$ in the low-frequency limit.
This can be seen as follows.

Consider the limit $\omega\to 0$ of $\Psi_{\omega 0} \equiv \psi_{\omega 0}/A^{\text{tr}}_{\omega 0}$, where $A^{\text{tr}}_{\omega l}$ is defined by Eq.~(\ref{asy_hor}).
We have $\Psi_{\omega 0} \to 1$ near the black hole event horizon.
Near the cosmological horizon we have instead
\begin{equation}
\lim_{\omega \to 0}\Psi_{\omega 0} = \lim_{\omega \to 0}\left(\frac{A^{\text{in}}_{\omega 0}}{A^{\text{tr}}_{\omega 0}}e^{-i\omega r_*}
+ \frac{A^{\text{out}}_{\omega 0}}{A^{\text{tr}}_{\omega 0}}e^{+i\omega r_*}\right).
\end{equation}
Now, if there is a solution to Eq.~(\ref{radial_eq}) for $l=0$ and $\omega=0$ that tends to $1$ as $r_*\to -\infty$ and to a constant $a$ as $r_*\to +\infty$, then we must have
\begin{equation}
a = \lim_{\omega \to 0}\left(\frac{A^{\text{in}}_{\omega 0}}{A^{\text{tr}}_{\omega 0}}+\frac{A^{\text{out}}_{\omega 0}}{A^{\text{tr}}_{\omega 0}}\right).
\end{equation}
Then the $\omega\to 0$ limit of $\gamma_0(\omega) = |A^{\text{tr}}_{\omega 0}/A^{\text{in}}_{\omega 0}|^2$ must be nonzero.
This is indeed what happens for $\xi = 0$~\cite{BCKL}.
However, generically the $\omega=0$ solution which tends to $1$ as $r_*\to -\infty$ behaves as $\Psi_{0 0} \approx s\,r_*$ in the limit $r_*\to +\infty$, where $s$ is a constant.
In such cases $A^{\text{out}}_{\omega 0} \to - A^{\text{in}}_{\omega 0}$ in the limit $\omega\to 0$ and we have
\begin{equation}
s = -2i\lim_{\omega \to 0}\omega\frac{A^{\text{in}}_{\omega 0}}{A^{\text{tr}}_{\omega 0}}\,.
\end{equation}
This leads to the following small $\omega$ approximation of the transmission coefficient:
\begin{equation}
\gamma_{0}(\omega)=\left|\frac{A^{\text{tr}}_{\omega 0}}{A^{\text{in}}_{\omega 0}}\right|^2 \approx \frac{4}{s^2}\omega^2 \label{btogamma}
\end{equation}
as in the asymptotically-flat case.
In the rest of this subsection we calculate the constant $s$ for small $\xi$ to second order in $\xi$ and find the low-frequency behavior of $\gamma_0(\omega)$ to this order.

We first write Eq.~(\ref{radial_eq}) with $\omega=0$ in terms of $R(r)\equiv \Psi_{0 0}(r)/r$ as
\begin{equation}
\frac{d\ }{dr}\left[ f(r)r^2\frac{dR}{dr}\right] = 4\xi\Lambda r^2R(r). \label{Requation}
\end{equation}
We take $R^{(0)}(r)= 1/r_H$, i.e.\ $\Psi_{00} = r/r_H$, as the lowest-order solution.
Then by writing
\begin{equation}
R(r) = 1/r_H + \xi R^{(1)}(r) + O(\xi^2)
\end{equation}
we have
\begin{equation}
\frac{d\ }{dr}\left[ f(r)r^2\frac{dR^{(1)}}{dr}\right] = \frac{4 \Lambda}{r_H} r^2.
\end{equation}
This equation can readily be solved by integration.
We find
\begin{eqnarray}
R^{(1)}(r) & = & - \frac{4 (r_C^2+r_Cr_H+r_H^2)}{r_Cr_H(2r_C+r_H)}\log\left(\frac{r_C-r}{r_C-r_H}\right) \nonumber \\
&& - \frac{4(r_C^2+r_Cr_H+r_H^2)}{r_H(r_C+r_H)(2r_C+r_H)}\log\left(\frac{r+r_C+r_H}{r_C+2r_H}\right)\nonumber \\
&& + \frac{4 r_H}{r_C(r_C+r_H)}\log\frac{r}{r_H}. \label{seed2}
\end{eqnarray}

Next we note that Eq.~(\ref{Requation}) can be written as
\begin{equation}
\frac{d\ }{dr}\left[ r\frac{d\Psi_{00}}{dr_*} - f(r)\Psi_{00}\right] = 4\xi \Lambda r^2R(r).
\end{equation}
Since $f(r)\to 0$ like $r_C-r$ as $r\to r_C$ whereas from~\eqref{tortoise} we see that $\Psi_{00}$ diverges like $\ln(r_C-r)$ in this limit, we conclude that $f(r)\Psi_{00}\to 0$ as $r\to r_C$.
Recalling that we have required $d\Psi_{00}/dr_*\to 0$ as $r\to r_H$ and that we have defined $\lim_{r\to r_C}d\Psi_{00}/dr_* = b$, we find
\begin{equation}
b = \frac{4\xi \Lambda}{r_C} \int_{r_H}^{r_C} r^2R(r)dr.
\end{equation}
By substituting $R(r)=1/r_H + \xi R^{(1)}(r)$, where $\xi R^{(1)}(r)$ is given by Eq.~(\ref{seed2}) into this equation we find the constant $b$ to order $\xi^2$ as
\begin{equation}
b = \frac{4\xi(r_C-r_H)}{r_Cr_H}\left[1+4\xi G(r_C,r_H)\right], \label{an_approx}
\end{equation}
where
\begin{eqnarray}
G(r_C,r_H) &  = & \frac{4}{3} + \frac{r_C^2r_H^2}{(r_C+r_H)(r_C^3-r_H^3)}\log \frac{r_C}{r_H}\nonumber \\
&&  - \frac{r_C^2+r_Cr_H+r_H^2}{r_C^2-r_H^2}\log \frac{2r_C+r_H}{r_C+2r_H}.
\end{eqnarray}
Hence by Eq.~(\ref{btogamma})
\begin{equation}
\gamma_0(\omega) \approx \frac{r_C^2r_H^2}{4\xi^2(r_C-r_H)^2} \,\, \omega^2 \,\,
\left[ 1+ 4\xi G(r_C,r_H)\right]^{-2}
\label{g0_ana}
\end{equation}
to lowest order in $\xi$.

\subsection{Low-frequency approximation with small $\lambda$}
\label{lowfreq}

In this section we present an analytic computation that provides an approximation to the greybody factor of small dS black holes in the low-frequency regime.
The calculation is performed using matched asymptotic expansions, a technique first employed in this context in Ref.~\cite{Starobinsky:1973}.
Our calculation follows along the same lines as Refs.~\cite{Unruh:1976, Cardoso:2004hs}. 

The main interest of this result is that it is valid for arbitrary angular quantum number $l$ and coupling $\xi$.
On the other hand, the accuracy of the result is guaranteed only if the two asymptotic regions significantly overlap, which implies that it is only valid for small frequencies.
In addition, the procedure is justified, as we shall see, only for the class of ``small'' black holes (compared with the characteristic dS scale), i.e. $\lambda \ll 1$. Thus the approximation for the greybody factor we shall obtain below, Eq.~\eqref{first-result}, and the result of the previous section, Eq.~\eqref{g0_ana}, are valid in complementary regions of parameter space.

The starting point is the radial wave equation~\eqref{radial_eq}, written in terms of $X_{\omega l}(r) \equiv \psi_{\omega l}(r)/r$, namely
\begin{eqnarray}
r^2f\frac{d}{dr}\left[ r^2f\frac{d}{dr}X_{\omega l} \right]
&-& r^2f \left[ l(l+1)+4\xi\Lambda r^2 \right]X_{\omega l} \nonumber\\
&+&\omega^2 r^4 X_{\omega l} = 0\,.
\end{eqnarray}

We then analyze this wave equation in two distinct, but overlapping, regions.
The {\em near} region is defined by $r-r_H\ll1/\omega$ whereas the {\em far} region is such that $r-r_H\gg 2M$.
The two regions overlap if $\omega M\ll 1$.

The essential point of considering small SdS black holes is that in the near region $r\sim r_H\sim 2M$ we can discard the effects of the cosmological constant.
Hence $r^2f(r) \approx r^2 -2Mr$ and the near region wave equation becomes
\begin{eqnarray}
(r^2-2Mr)\frac{d}{dr}\left[ (r^2-2Mr)\frac{d}{dr}X_{\omega l}^{(near)} \right] \qquad&\,& \nonumber\\
- \left[ (r^2-2Mr) l(l+1) - (2M)^4\omega^2 \right] X_{\omega l}^{(near)} &=& 0\,.\;\;
\end{eqnarray}
The (purely ingoing) solution of this equation can be written in terms of a hypergeometric function~\cite{Cardoso:2004hs}:
\begin{eqnarray}
X_{\omega l}^{(near)}(r) = A \left( 1-\frac{2M}{r} \right)^{-2iM\omega} \left(\frac{2M}{r}\right)^{l+1} \quad&\,&  \nonumber\\
\times F\left(l+1, l+1-4iM\omega, 1-4iM\omega\, ; 1-\frac{2M}{r}\right)&,&\;\;
\label{near_sol}
\end{eqnarray}
with $A$ being the constant amplitude.
It can be shown that
\begin{equation}
X_{0l}^{(near)}(r) = (-1)^l A P_l\left(1-\frac{r}{M}\right),
\end{equation}
where $P_l(x)$ are the Legendre polynomials~\cite{Candelas80}.

For the far region solution we can neglect the effects of the black hole, so $M\sim0$ and $f(r)\approx 1-\Lambda r^2/3$.
Defining a new variable $x\equiv 1-\Lambda r^2/3$ the radial wave equation then becomes
\begin{eqnarray}
x(1&-&x)\frac{d^2 X_{\omega l}^{(far)}}{dx^2}+\left(1-\frac{5}{2}x\right)\frac{d X_{\omega l}^{(far)}}{dx} \nonumber\\
&-& \left[ \frac{l(l+1)}{4(1-x)} - \frac{3\omega^2}{4\Lambda x} + 3\xi \right] X_{\omega l}^{(far)} = 0\,,
\end{eqnarray}
whose general solution can again be expressed in terms of hypergeometric functions as
\begin{eqnarray}
\label{far_sol}
X_{\omega l}^{(far)}(x) = C x^{i\frac{\omega}{2}\sqrt{\frac{3}{\Lambda}}} \left(1-x\right)^{l/2} F(\nu_{+}, \nu_{-}, \mu\,; x) \qquad\qquad \\
+ D x^{-i\frac{\omega}{2}\sqrt{\frac{3}{\Lambda}}} \left(1-x\right)^{l/2} F(1+\nu_{+}-\mu, 1+\nu_{-}-\mu, 2-\mu\,; x)\,,  \nonumber
\end{eqnarray}
where
\begin{eqnarray}
\nu_{\pm} & = & \frac{1}{4}\left(3+2l \pm \sqrt{9-48\xi} + 2i\omega\sqrt{\frac{3}{\Lambda}}\right),\\
\mu & = & 1 + i\omega\sqrt{\frac{3}{\Lambda}}.
\end{eqnarray}
(See also Ref.~\cite{HiguchiCQG87}.)

Having the two asymptotic solutions at hand, one can now match them in the overlapping region $2M\ll r\ll 1/\omega$.
This requires finding the large-$r$ behavior of Eq.~\eqref{near_sol} and the small-$r$ limit of Eq.~\eqref{far_sol}, which corresponds to $x\to1$.
In the overlapping region both solutions are expressed as a superposition of two terms, with corresponding behaviors $\sim r^l$ and $\sim r^{-(l+1)}$.
By matching the respective coefficients we determine the constants $C$ and $D$ as functions of all the parameters in our problem, namely $\omega, \Lambda, M, l$ and $\xi$.
Obviously, both coefficients $C$ and $D$ are proportional to the amplitude $A$, so when one computes their ratio, $A$ cancels out.

The greybody factor is computed from Eq.~\eqref{gbf2} so we must relate coefficients $C$ and $D$ with $A_{\omega l}^{\rm in}$ and $A_{\omega l}^{\rm out}$.
The solution near the cosmological horizon, $x\sim 0$, behaves like
\begin{equation}
X_{\omega l} \approx C x^{i\frac{\omega}{2}\sqrt{\frac{3}{\Lambda}}} + D x^{-i\frac{\omega}{2}\sqrt{\frac{3}{\Lambda}}}\,.
\label{asy_sol}
\end{equation}
Also, in this spacetime region ($x\sim 0$) the coordinate $x$ is related with the tortoise coordinate $r_*$ through
\begin{equation}
r_*\approx -\frac{1}{2}\sqrt{\frac{3}{\Lambda}} \log\frac{x}{2}\,.
\end{equation}
Thus, comparing Eq.~\eqref{asy_sol} with Eq.~\eqref{psi_rc} we conclude that
\begin{equation}
C = \frac{A_{\omega l}^{\rm in}}{r_C}\, 2^{-i\frac{\omega}{2}\sqrt{\frac{3}{\Lambda}}}\,, \qquad
D = \frac{A_{\omega l}^{\rm out}}{r_C}\, 2^{i\frac{\omega}{2}\sqrt{\frac{3}{\Lambda}}}\,,
\end{equation}
and therefore the greybody factor may be expressed as
\begin{equation}
\gamma_l(\omega) = 1- \frac{|D|^2}{|C|^2}\,.
\end{equation}
Substituting into the above expression the coefficients $C$ and $D$ obtained from the matching procedure, we arrive at the following analytic small-frequency approximation for the greybody factor of small dS black holes:
\begin{eqnarray}
\gamma_l(\omega) &=& \frac{16 \pi^{7/2} (-1)^l l! \left((2l)!\right)^2  \Gamma(2l+2)\Gamma(-l-1/2) }{ -\cos\left(\pi\sqrt{9-48\xi}\right) - \cosh\left(2\pi\frac{\omega M}{\sqrt{\lambda}}\right)  } \nonumber\\
&& \times \frac{ \sinh\left(\pi\frac{\omega M}{\sqrt{\lambda}}\right) \lambda^{l+1/2} }{\sinh(4\pi\omega M)\, |H(\omega)|^2} \,,  \label{first-result}
\end{eqnarray}
where the function $H(\omega)$ has been defined as
\begin{eqnarray}
H(\omega) &=& (2l)! \Gamma(2l+2)\Gamma\left(l+\frac{1}{2}\right) \nonumber\\
&&\times \, \Gamma(-l-4i\omega M)\Gamma(\alpha_+)\Gamma(\alpha_-) \nonumber\\
&+& \, (-4)^l (l!)^2 \lambda^{l+\frac{1}{2}} \Gamma\left(-l-\frac{1}{2}\right) \nonumber\\ 
&& \times \Gamma(1+l-4i\omega M)\Gamma(\beta_+)\Gamma(\beta_-)\,,
\end{eqnarray}
and
\begin{eqnarray}
\alpha_{\pm} &=& \frac{1}{4}\left(1-2l \pm \sqrt{9-48\xi}-2i\frac{\omega M}{\sqrt{\lambda}}\right)\,,\\
\beta_\pm &=& \frac{1}{4}\left(3+2l \pm \sqrt{9-48\xi}-2i\frac{\omega M}{\sqrt{\lambda}}\right)\,.
\end{eqnarray}

This result is fully consistent with Refs.~\cite{BCKL, KGB, HNS_atmp14_727}, meaning that the zero-frequency limit of the greybody factor for $l=\xi=0$ reproduces Eq.~\eqref{gbfmc0} in the small SdS black hole regime.
For $\xi\neq 0$ Eq.~\eqref{first-result} becomes in the low-frequency limit
\begin{equation}
\gamma_l(\omega) = \frac{\pi(l!)^2\lambda^l \left|\Gamma(\beta^0_+)\Gamma(\beta^0_-)\right|^2}{4^l|\Gamma(l+\frac{1}{2})\Gamma(l+\frac{3}{2})|^2}(r_H\omega)^2, \label{simplified}
\end{equation}
where $\beta^0_\pm = \beta_{\pm}|_{\omega=0}$. This result reproduces Eq.~\eqref{g0_ana} for $l=0$ with small $\xi$ at zeroth order in $\lambda$.
Moreover, it is applicable to more general cases, with $\xi \neq 0$ (not necessarily small) and general $l$.

\subsection{Computational methods and numerical results}
\label{gbfnr}

In this section we present results valid for arbitrary
frequencies that were obtained by solving numerically
Eq.~(\ref{radial_eq}).
We develop this solution from $r = r_H(1 + \varepsilon_1)$ to $r = r_C(1 - \varepsilon_2)$, with $\varepsilon_1,\varepsilon_2 \ll 1$.
The reflection coefficient $|A_{\omega l}^{\text{out}}/A_{\omega l}^{\text{in}}|^2$ may be obtained by comparing the numerical solution with the asymptotic form~(\ref{psi_rc}).

\begin{figure}
\centering
\includegraphics[width=8.6cm]{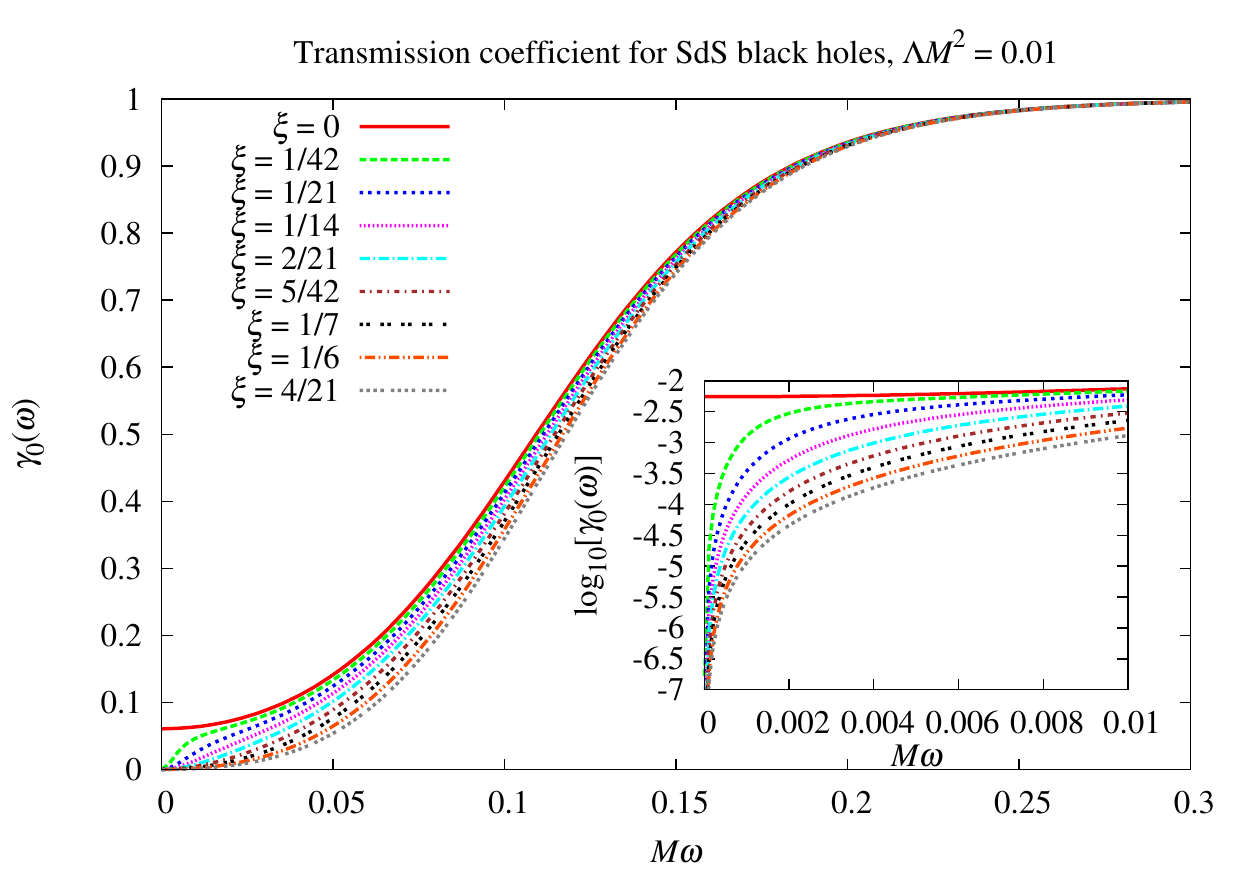}
\caption{Greybody factor plotted as a function of the frequency
for $\Lambda M^2 = 0.01$ and
for different choices of the coupling $\xi$.}
\label{ngbf}
\end{figure}

In FIG.~\ref{ngbf} we plot our numerical results for the greybody factor.
We see that the results go to zero as $\omega \to 0$, except for the minimally coupled case.
The behavior of the transmission coefficient in the SdS spacetime (except for the minimally
coupled case) keeps the same structure as in asymptotically flat
spacetimes: it is zero in the low-frequency limit and tends to unity
in the high-frequency limit. The explanation for this behavior in
asymptotically flat spacetimes relies on the semiclassical approximation,
in which we can relate the apparent impact parameter with the wave
frequency and angular momentum as $b \approx (l+1/2)/\omega$. Roughly
speaking, for a fixed $l$, if $b > b_c$, where $b_c$ is the critical
impact parameter~\cite{Wald}, the transmission coefficient is zero.
As $\omega$ increases, $b$ approaches the value of $b_c$, and when
$b < b_c$ the particle is absorbed by the black hole. In
Sec.~\ref{analytic_results-abs} we develop the geodesic analysis in the SdS spacetime,
and, although the concepts of impact parameter are physically different
in asymptotically flat and dS spacetimes, this analysis
may be helpful in understanding the absorption process in asymptotically dS
spacetimes.

\begin{figure*}
\centering
\includegraphics[width=17.2cm]{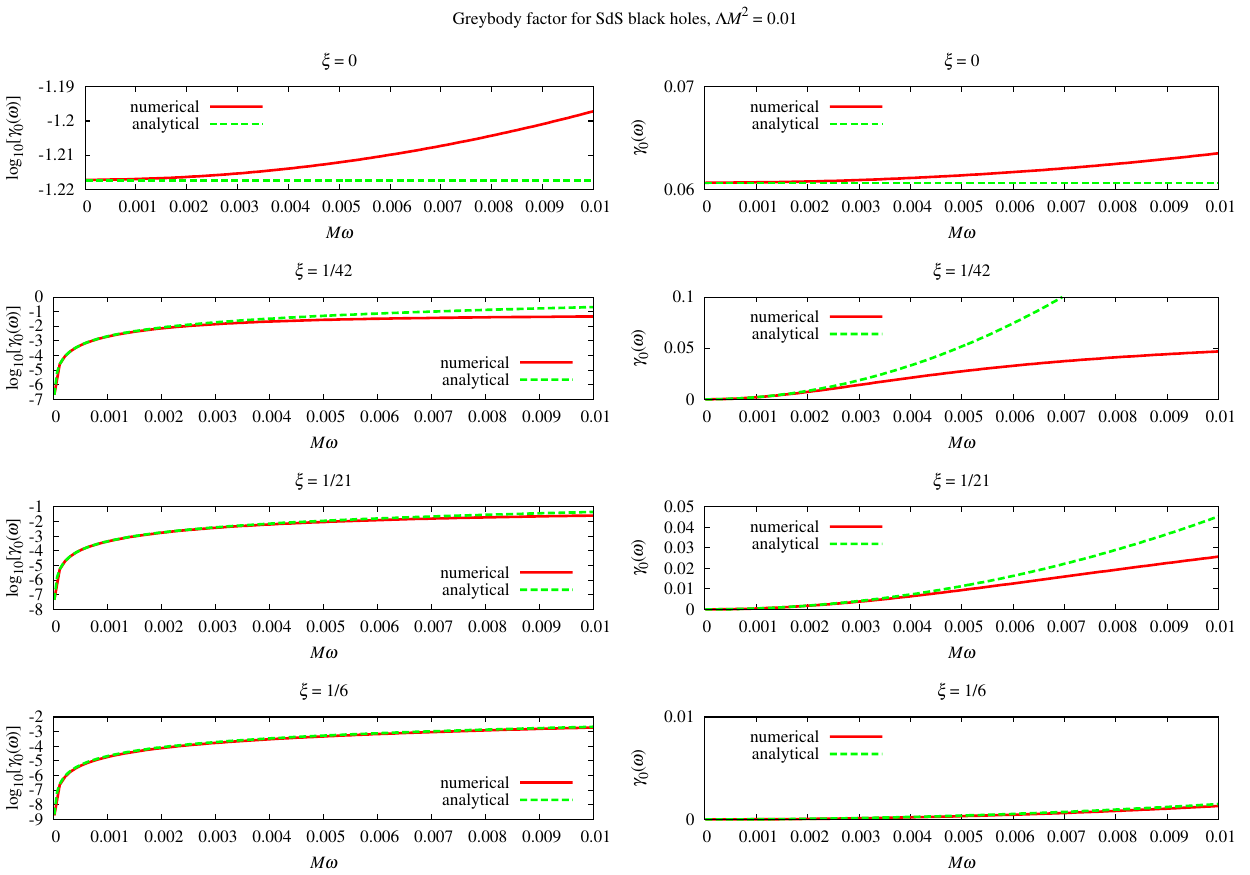}
\caption{Greybody factor plotted as a function of the frequency
for $\Lambda M^2 = 0.01$ compared with the analytic approximation 
given by Eq.~\eqref{g0_ana},
for different choices of the coupling $\xi$.
Note the logarithmic scale on the vertical axis of the left plots.}
\label{gbf_an}
\end{figure*}

In FIG.~\ref{gbf_an} we compare the numerical results
for the greybody factor with the ones obtained using the analytical
approximation~(\ref{g0_ana}). We find good agreement for low $\omega$ even
in the conformally coupled case ($\xi = 1/6$).
(We recall that Eq.~(\ref{g0_ana}) is valid in the small-$\xi$ regime.)

\begin{figure*}
\centering
\subfigure{\includegraphics[width=8.9cm]{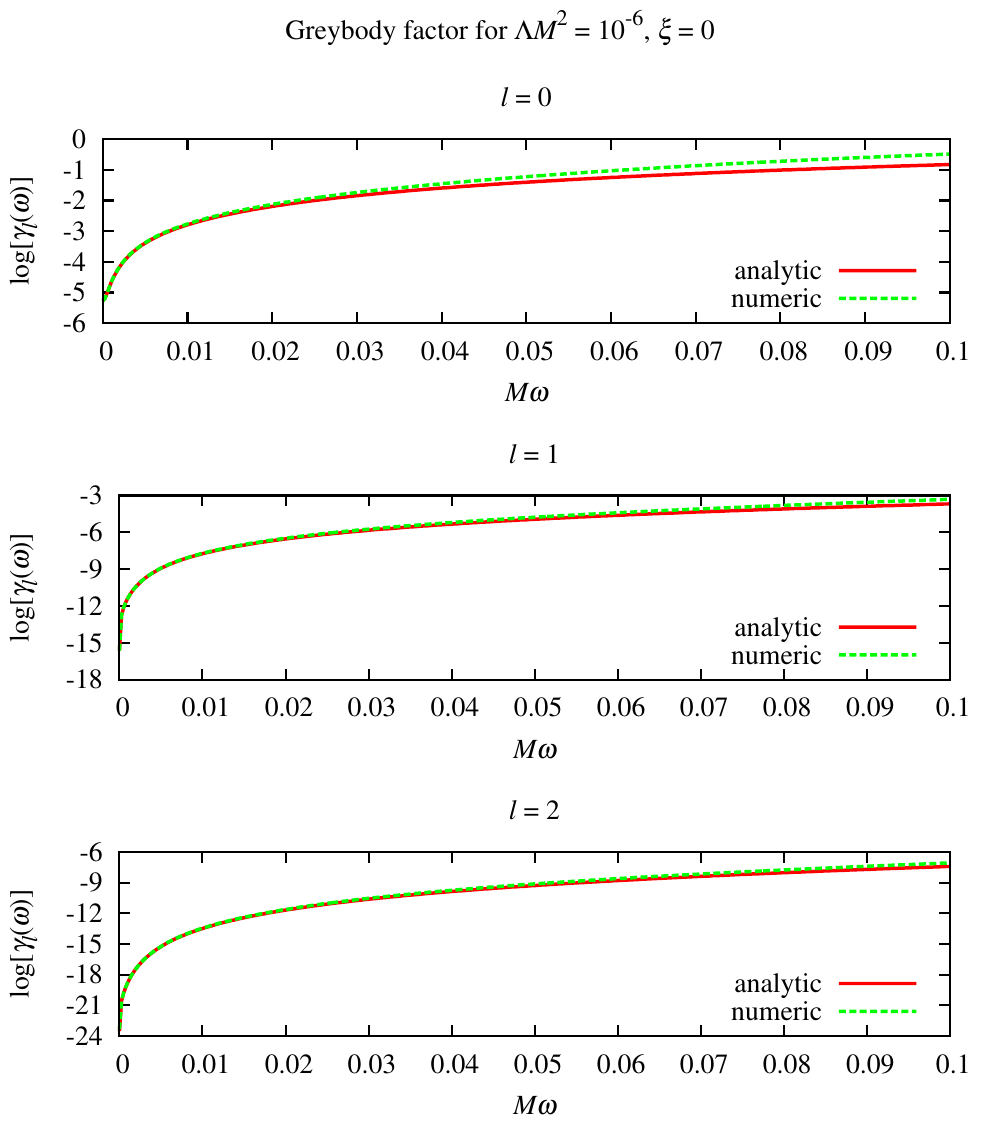}\label{sbh:a}}
\subfigure{\includegraphics[width=8.9cm]{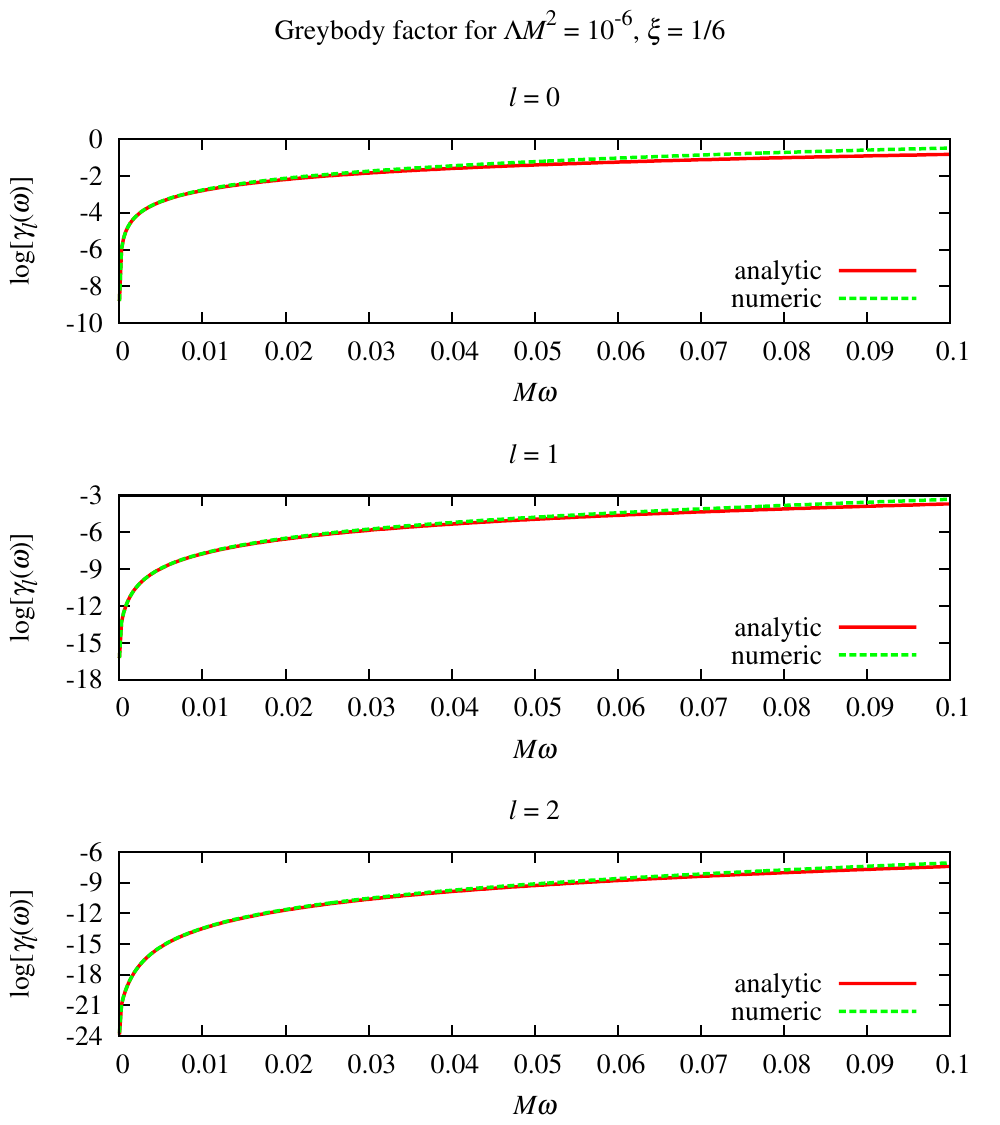}\label{sbh:b}}
\caption{\emph{Left plots:} Greybody factor for small black holes in which
$\Lambda M^2 = 10^{-6}$,  $\xi = 0$, and $l=0,1,2$.
Note the logarithmic scale on the vertical axes.
\emph{Right plots:} Analogous plots for $\xi = 1/6$. 
Excellent agreement is found
between the numerical results and the analytic results given by
Eq.~\eqref{first-result}.}
\label{sbh}
\end{figure*}

In FIG.~\ref{sbh} numerical results are compared with the analytic
results for small black holes, given by Eq.~\eqref{first-result}, for
$\Lambda M^2 = 10^{-6}$, $l=0,1,2$, with $\xi = 0$ (left plots), and
$\xi = 1/6$ (right plots). We note an excellent agreement
in the low-frequency limit, which is the regime of validity of
approximation~\eqref{first-result}. This can be regarded as a simple
consistency check of our results.

\section{Energy emission}
\label{ee}

The number of massless scalar particles emitted by the black hole per unit time,
also called flux spectrum, is given by
\begin{equation}
\frac{d N (\omega)}{dt} =
\frac{d\omega}{2\pi} \,
\frac{1}{e^{\omega / T_H} - 1}\,
\sum_{l=0}^{\infty} \left( 2l+1 \right) \gamma_l(\omega) .
\label{fs}
\end{equation}
The differential energy emission rate reads
\begin{equation}
\frac{d^2 E (\omega)}{dt d\omega} =
\frac{1}{2\pi}
\frac{\omega}{e^{\omega / T_H} - 1}\,
\sum_{l=0}^{\infty} \left( 2l+1 \right) \gamma_l(\omega) \,.
\label{deer}
\end{equation}

\begin{figure}
 \includegraphics[width=8.6cm]{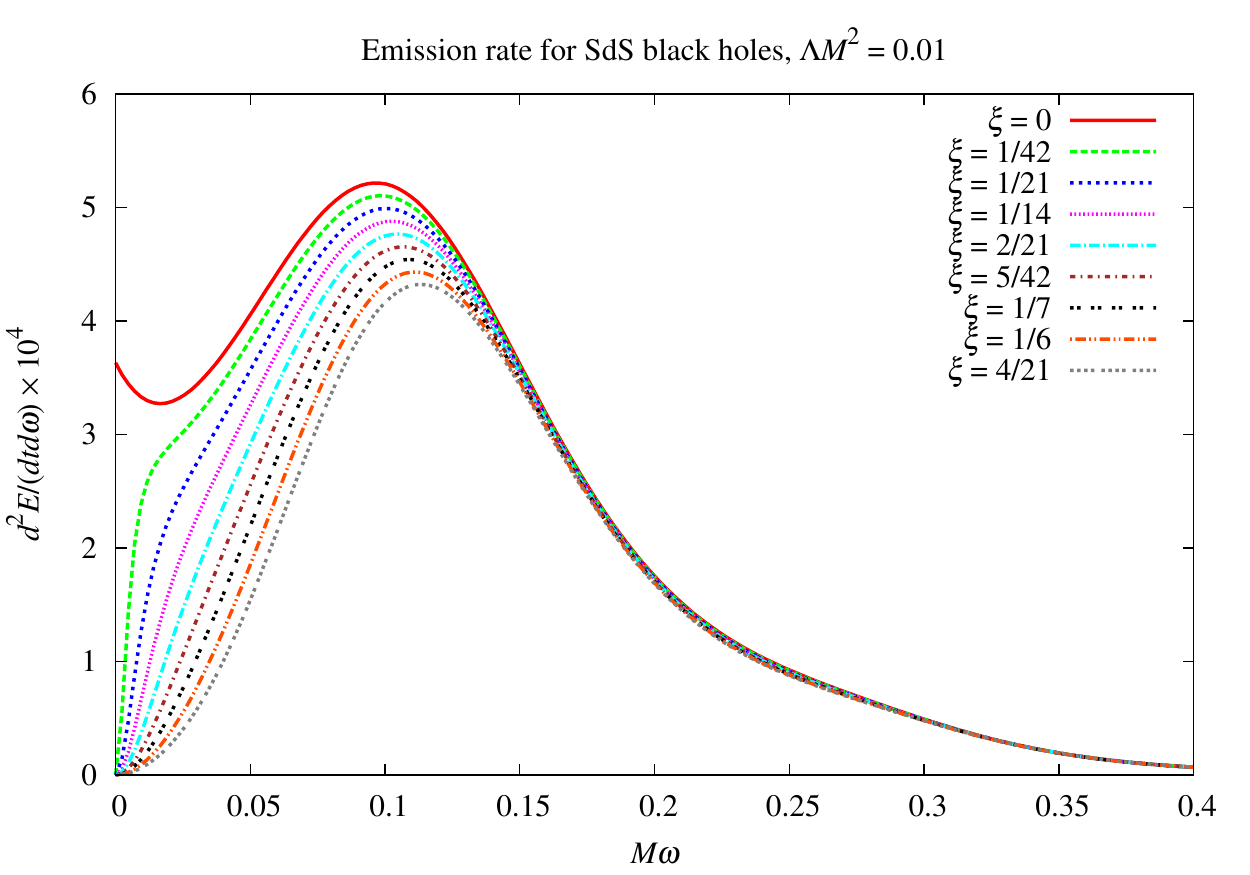}
 \includegraphics[width=8.6cm]{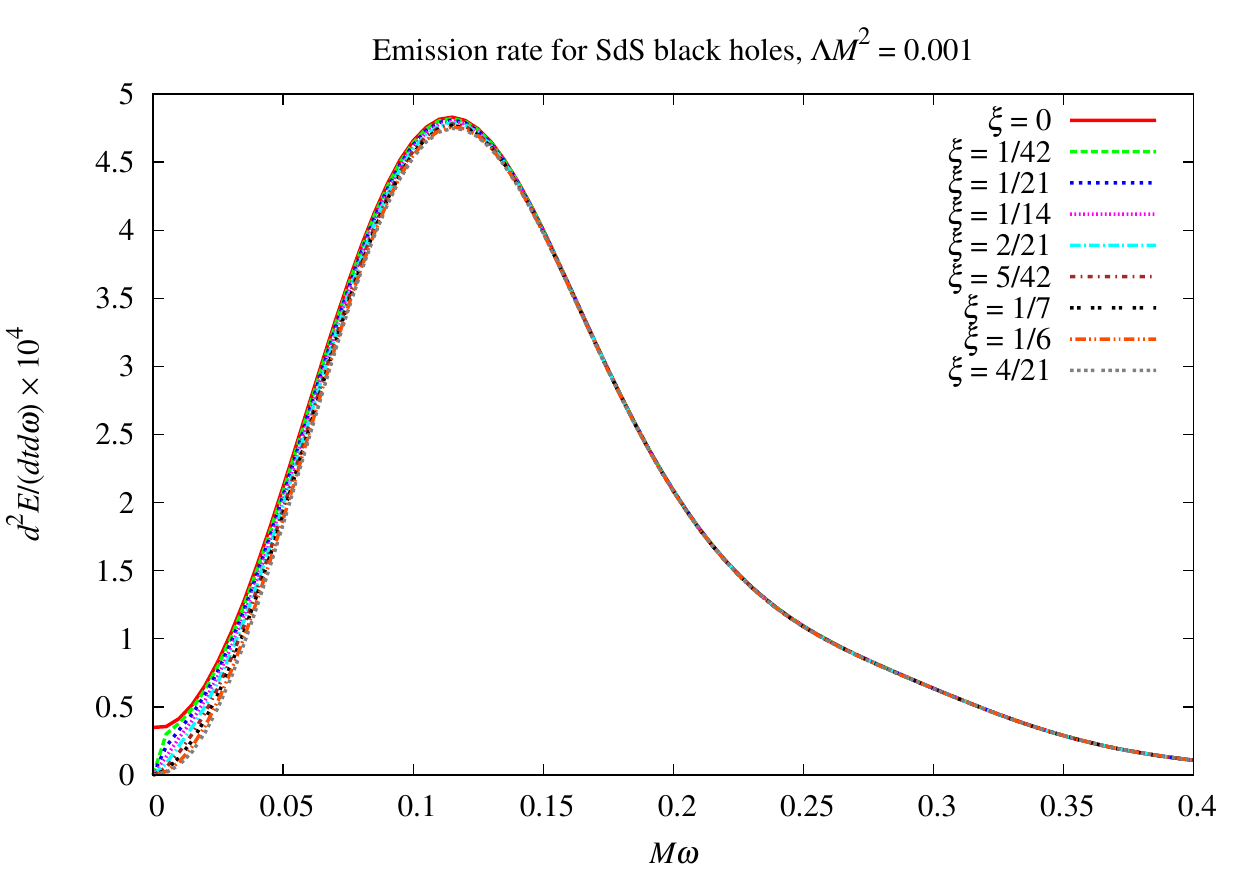}
 \caption{Emission rate for Schwarzschild de-Sitter black holes for
 $\Lambda M^2 = 0.01$ (top panel) and $\Lambda M^2 = 0.001$ (bottom panel), and different values of $\xi$.
 A non-vanishing emission rate at zero frequency occurs only in the minimally coupled case ($\xi = 0$).}
 \label{emission_rate}
\end{figure}

In FIG.~\ref{emission_rate} we plot ${d^2 E (\omega)}/{(dt d\omega)}$.
As for the transmission coefficient, the emission rate for zero frequency is nonzero only in the
case of a minimally coupled scalar field. 
Therefore, we conclude that the conjectured distinctive feature
caused by the presence of the non-vanishing cosmological constant,
namely the emission of a significant number of ultrasoft quanta,
pointed out in Ref.~\cite{KGB} (even in the four-dimensional case --
cf. Section V of that paper), will not occur for the case of a
non-minimally coupled scalar field.
Note, however, that there is still an enhancement in the emission of soft quanta for low 
$\xi$ with the rate dropping to zero rapidly as $\omega$ becomes very close to zero.

It is apparent in FIG.~\ref{emission_rate} that the difference in the emission rates for distinct couplings is only considerable in the low-frequency regime. In other words, the coupling to the scalar curvature becomes irrelevant above intermediate values of the frequency ($\omega \sim M^{-1}$).
We also note that these differences for low frequencies become more and more pronounced as the value of
$\lambda$ increases.
On the other hand, the enhancement in the emission rate occurring at low frequencies becomes less significant as the coupling parameter $\xi$ grows.

\section{Generalized absorption cross section}
\label{acs}

The definition of absorption cross section in non-asymptotically flat spacetimes cannot be formulated as it is done in asymptotically flat geometries.
In the latter case, one may consider an impinging plane wave, $\Phi_{\text{inc}} = e^{i\omega (z - t)}$, which is a solution of the field equation in the asymptotic region since the metric there approaches that of Minkowski spacetime.
After being scattered off the spherically symmetric black hole, the scalar field in the asymptotic region $r \to \infty$ can be decomposed into spherical harmonics~\cite{prd76_107502}
\begin{eqnarray}
 \Phi_{\text{sca}} & \approx & i \sum\limits_{l = 0}^{\infty} \frac{(-1)^l (2l+1)}{2\omega}
 \frac{ \left[ \ e^{-i\omega r} + R_{\omega l}e^{i\omega r} \right] }{r} \nonumber \\
&&  \times P_l (\cos\theta) e^{-i\omega t},
\label{flat-sca_wave}
\end{eqnarray}
with $|R_{\omega l}|^2$ being the reflection coefficient.
Then, using the definition of the absorption cross section~\cite{Cohen}
\begin{equation}
\sigma \equiv \frac{\text{absorbed flux}}
{\text{incident wave current}}\,,
\label{sigma}
\end{equation}
one concludes that the total absorption cross section for an asymptotic plane wave incident on an asymptotically flat spherically symmetric spacetime is
\begin{equation}
\sigma =
\sum^{\infty}_{l=0} \sigma_l =
\frac{\pi}{\omega^{2}}
\sum^{\infty}_{l=0}
{ \left( 2l+1 \right)
\gamma_l(\omega) }\,,
\label{tacs}
\end{equation}
where $\sigma_l$ represents the absorption cross section of each partial wave~\cite{Sanchez}, usually referred to as partial absorption cross section.

In the case of asymptotically dS spacetimes it is easy to see that a plane wave is no longer an asymptotic solution of the field equation.
Indeed, $\psi_{\omega l}(r)=e^{\pm i\omega r}$ does not satisfy Eq.~\eqref{radial_eq} near the cosmological horizon $r_C$, and therefore a sum over angular modes like the one displayed on the right hand side of Eq.~\eqref{flat-sca_wave} does not yield an asymptotic solution.
However, $\psi_{\omega l}(r)=e^{\pm i\omega r_*}$ is an asymptotic solution of the Regge-Wheeler equation~\eqref{radial_eq} and one might try to construct an analogue of a plane wave out of such solutions.

Let us consider then an incident wave of the form
\begin{equation}
\Phi_{\text{inc}}^{\text{dS}} = \frac{r_*}{r} e^{i\omega (r_*\cos\theta-t)}\,,
\label{dSincident}
\end{equation}
which can also be decomposed into spherical waves.
This is understood by first recalling the identity~\cite{Gradshteyn}
\begin{equation}
 e^{i\omega r_* \cos\theta} = \sum\limits_{l=0}^{\infty} i^l (2l+1)
 j_l (\omega r_*) P_l(\cos\theta)\,,
 \label{plane_wave}
\end{equation}
and then using the asymptotic form of the spherical Bessel functions
of the first kind, $j_l(x)$, thus obtaining
\begin{eqnarray}
 \frac{r_*}{r} e^{i\omega r_* \cos\theta} & \approx & i \sum\limits_{l =0 }^{\infty}
 \frac{(-1)^l(2l+1)}{2\omega} \frac{\left[ e^{-i\omega r_*}- (-1)^l e^{i\omega r_*} \right] }{r} \nonumber \\
 & & \times P_l(\cos\theta)\,.
 \label{dS_wave}
\end{eqnarray}
This approximation is valid as an expansion for $\omega r_* \gg 1$.

It turns out that the wave~\eqref{dSincident} is {\em not} an asymptotic solution of the Klein-Gordon equation~\eqref{KG}. Accordingly, $\Phi=r^{-1}e^{-i\omega(t\mp r_*)}Y_{lm}(\theta,\varphi)$ is also not an asymptotic solution of~\eqref{KG}, even though it is an asymptotic solution of the Regge-Wheeler equation~\eqref{radial_eq}~\cite{footnote2}.

Nevertheless, for small black holes there exists an intermediate region $r_H \ll r \ll r_C$ where $ f \approx 1$ and $ r_* \approx r$.
In this region the wave~\eqref{dSincident} is an {\em approximate} solution of the field equation, meaning that the Klein-Gordon operator defined by the left hand side of Eq.~\eqref{KG} does not annihilate~\eqref{dSincident} but instead yields terms that are suppressed by powers of $r_H/r$ and of $r/r_C$.

Let us now return to expression~\eqref{dS_wave}, which represents a scattered wave in pure dS spacetime at any  radial coordinate $r$ that is large compared to the wavelength $1/\omega$ but small relative to the cosmological horizon $r_C$.
In the presence of a black hole, the scattered wave will be modified by the inclusion of a non-trivial reflection
coefficient,
\begin{eqnarray}
 \Phi_{\text{sca}}^{\text{dS}} &\approx& i \sum\limits_{l =0 }^{\infty}
 \frac{(-1)^l(2l+1)}{2\omega} \frac{\left[ e^{-i\omega r_*} + R_{\omega l}
 e^{i\omega r_*} \right] }{r} \nonumber\\
  & & \times P_l(\cos\theta)e^{-i\omega t}\,,
 \label{dS_sca}
\end{eqnarray}
with $R_{\omega l} = A_{\omega l}^{\text{out}}/A_{\omega l}^{\text{in}}$ and $A_{\omega l}^{\text{in}} = i(-1)^l\sqrt{\pi(2l+1)}/\omega$.
Despite the similarity between Eqs.~\eqref{dS_sca} and~\eqref{flat-sca_wave}, in order to define the absorption cross section one must still determine the flux of the wave~\eqref{dS_sca} at some intermediate radius $r\in (r_H , r_C)$ and the current of the incident wave.

The wave flux is given by
\begin{equation}
 \mathcal{F} = - \int J^r r^2 d\Omega\,,
 \label{flux}
\end{equation}
where $d\Omega$ is the solid angle element and $J^r$ is the radial (contravariant) component of the current, which is defined for a wave $\Phi$ as
\begin{equation}
 J_\mu \equiv \frac{1}{2i}(\Phi^{*}\partial_\mu \Phi -
 \Phi\partial_\mu \Phi^{*})\,.
 \label{current}
\end{equation}
Using Eqs.~\eqref{dS_sca},~\eqref{flux}, and~\eqref{current}, it can be shown that the flux computed using the scattered wave is
\begin{equation}
 \mathcal{F} = \frac{\pi}{\omega} \sum_{l = 0}^{\infty} (2l+1)\gamma_l (\omega)\,,
 \label{sca_flux}
\end{equation}
which is independent of the radial coordinate $r$. To find Eq.~\eqref{sca_flux} we have used $\gamma_l (\omega) = 1 - |R_{\omega l}|^2$.

The incident current corresponding to wave~\eqref{dSincident} is found to be
\begin{equation}
 |\vec{J}_{\text{inc}}| = \omega \frac{r_*^2}{r^2}\left(f^{-1}\cos^2 \theta +
 \frac{r_*^2}{r^2}\sin^2\theta \right)^{1/2}.
 \label{inc_current}
\end{equation}
This result is general and could be applied for both asymptotically dS and flat spacetimes.
In the latter case, we may take the limit $r\to \infty$, in which case $r_* \to r$ and $f \to 1$, so that $|\vec{J}_{\text{inc}}| = \omega$.
This, together with Eq.~\eqref{sca_flux} and the definition of the absorption cross section~\eqref{sigma}, results in Eq.~\eqref{tacs}.

On the other hand, for asymptotically dS spacetimes if we take the limit $r\to r_C$ then $r_* \to \infty$ and $f \to 0$, which makes the incident current diverge. This would lead to an ill-defined absorption cross section and it is a consequence of the wave~\eqref{dSincident} not being a true asymptotic state.

However, for the case of small SdS black holes there exists an intermediate region where $ f \approx 1$ and $ r_* \approx r$, and therefore $|\vec{J}_{\text{inc}}| \approx \omega $.
Hence, it is physically sensible to define an absorption cross section only for small SdS black holes, i.e., for $\lambda \ll 1$. For this reason we refer to the quantity $\sigma$ in Eq.~\eqref{tacs} when used in asymptotically dS spacetimes as the ``generalized'' absorption cross section, and to the quantity $\sigma_l$ as the partial ``generalized'' absorption cross section. (We note that $\sigma$ in Eq.~\eqref{tacs} was referred to as absorption cross section in Ref.~\cite{KGB} for asymptotically dS spacetimes without explaining when this notion is physically meaningful.)

\subsection{Analytic results}
\label{analytic_results-abs}

Our numerical results show that,
in the low-frequency regime, the partial generalized
absorption cross section
is nonzero for all $l$ modes. This contrasts with
the case of the scalar field in
Schwarzschild spacetimes,
in which the greybody factor in the low-frequency regime behaves
generally as $\omega^{2l+2}$, and the partial absorption cross section
as $\omega^{2l}$ (see, for instance, Ref.~\cite{Oliveira}). 
Here, as shown in Eq.~\eqref{simplified}, the greybody factors behave
generally as $\omega^2$, except for the minimally coupled case,
where the greybody factor is constant in the low-frequency regime, as
expressed in Eq.~\eqref{gbfmc0}.

Although the concept of absorption cross section 
cannot be straightforwardly adopted in
asymptotically dS spacetimes, it is interesting that numerical results for
expression~\eqref{tacs} are consistent with what one could call the
``high-frequency generalized absorption cross section'', as we show here.

Let us present a geodesic analysis to illustrate how
this quantity can be found.
We start by recalling that the motion of a massless particle in the 
spacetime defined by the line element~\eqref{ds} is governed by:
\begin{equation}
 - f\dot{t}^2 + f^{-1}\dot{r}^2 + r^2 \dot{\phi}^2 = 0,
 \label{geo}
\end{equation}
where we have assumed  $\theta =\pi/2$,
without loss of generality
 (taking advantage of the spherical symmetry), 
 and the overdot represents the derivative with respect to an
affine parameter. The constants of motion are
\begin{eqnarray}
 E & \equiv & f \dot{t}, \\
 L & \equiv & r^2 \dot{\phi}.
 \label{geo_consts}
\end{eqnarray}
Writing Eq.~\eqref{geo} in terms of these constants, we get:
\begin{equation}
 \dot{r}^2 + \frac{fL^2}{r^2} = E^2,
 \label{geo2}
\end{equation}
which is similar to the equation for the total mechanical energy. 
In this analogy, the second term on the left hand side of Eq.~\eqref{geo2} 
plays the role of an ``effective potential'' for the
particle's motion of total energy $E^2$. This effective potential possesses
a maximum at $r = 3 M$, which implies that critical orbits exist in SdS spactime
at radius $r_c = 3 M$. This result does not depend on
the value of the cosmological constant and, therefore, is the same for
Schwarzschild and SdS spacetimes.

Making use of the analogy with asymptotically flat spacetimes once again,
we can define the quantity $b = L/E$, which is an analog of the apparent
impact parameter~\cite{Wald, Stuchlik:1999qk}. 
This quantity $b$,
which is related to the initial conditions of the particle motion,
has a critical value $b_c$ when
$E^2$ equals the maximum of the effective potential, given by
\begin{equation}
 b_c = 3M \left( \frac{1}{3} - 3 \Lambda M^2 \right)^{-1/2}.
 \label{bc}
\end{equation}
Classically, if $b < b_c$, then the particle is absorbed by the black
hole; if $b > b_c$, the particle is scattered
away from the black hole;
if $b = b_c$, the particle ends on the unstable orbit at $r = r_c$,
circling the black hole an infinite number of times.

We may push this analogy further 
and define what would be the ``high-frequency
generalized absorption cross section'' as:
\begin{equation}
 \sigma_{hf} \equiv \pi b_c^2 = 9\pi M^2
 \left( \frac{1}{3} - 3 \Lambda M^2 \right)^{-1} .
 \label{hfacs}
\end{equation}
We note that $\sigma_{hf} \to 27 \pi M^2$ as $\Lambda \to 0$, 
which is a well known result for Schwarzschild black holes. In the
next subsection we compare $\sigma_{hf}$ with the numerical results for
the quantity~\eqref{tacs} 
in the high-frequency limit
and obtain an excellent agreement.

\subsection{Numerical results}
\label{acsnr}

\begin{figure}
\centering
\includegraphics[width=8.6cm]{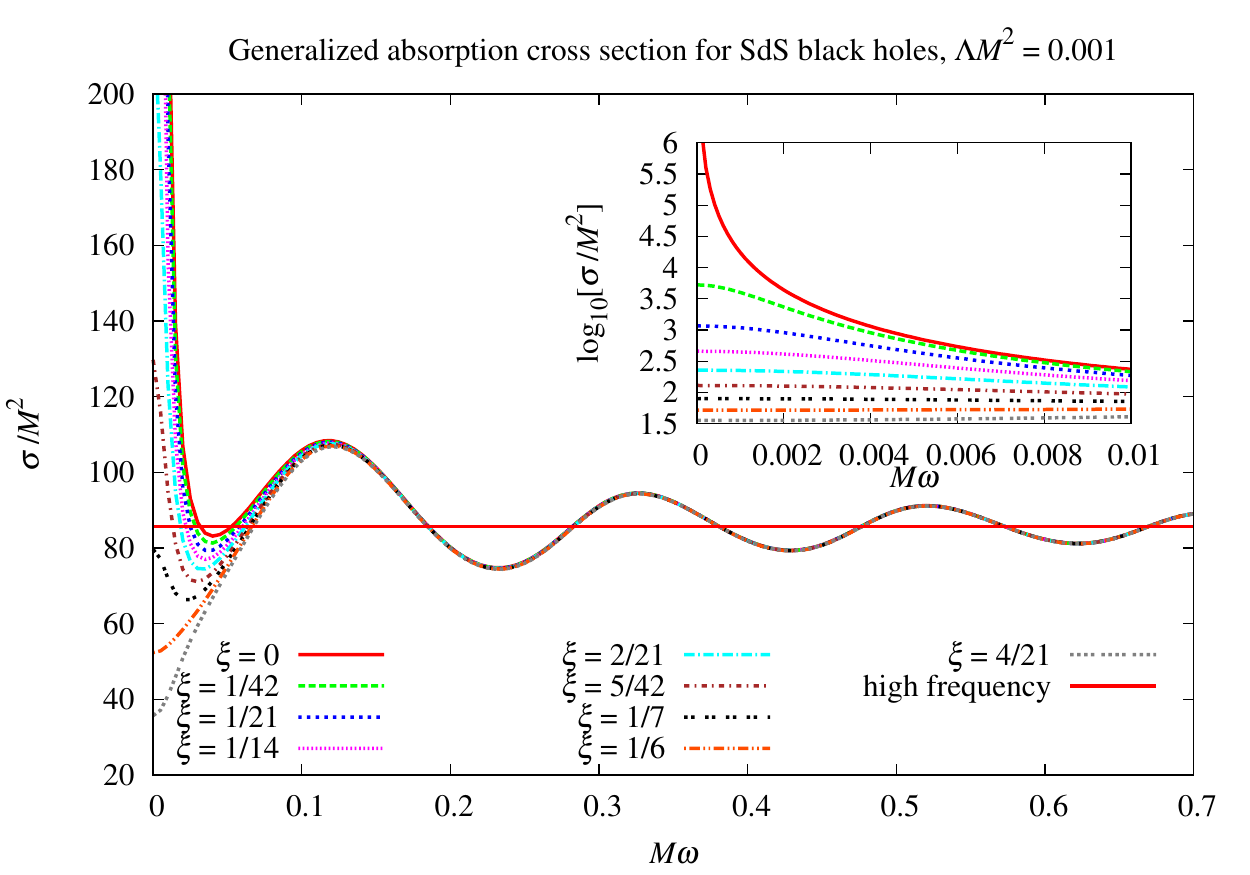}
\caption{Generalized absorption cross section plotted as a function of the
frequency for $\Lambda M^2 = 0.001$ and for different choices of the
coupling $\xi$. The summation in $l$ has been performed up to $l=3$.
The straight line is the high-frequency limit,  given by Eq.~\eqref{hfacs}.}
\label{nacs}
\end{figure}

In FIG.~\ref{nacs} we plot our numerical results for the 
generalized absorption cross section $\sigma$, defined by~\eqref{tacs}.
It diverges when $\omega \to 0$ in the minimally coupled case, but remains finite for
$\xi \neq 0$, as we can see from the log plots in the same figure.
The results presented in FIG.~\ref{nacs} agree very well with the
high-frequency limit given in Eq.~\eqref{hfacs}, which is
represented by the straight line, despite the absence of a 
clear physical meaning for the quantity $\sigma$ in generic
asymptotically dS spacetimes.

\begin{figure}
 \centering
 \includegraphics[width=8.6cm]{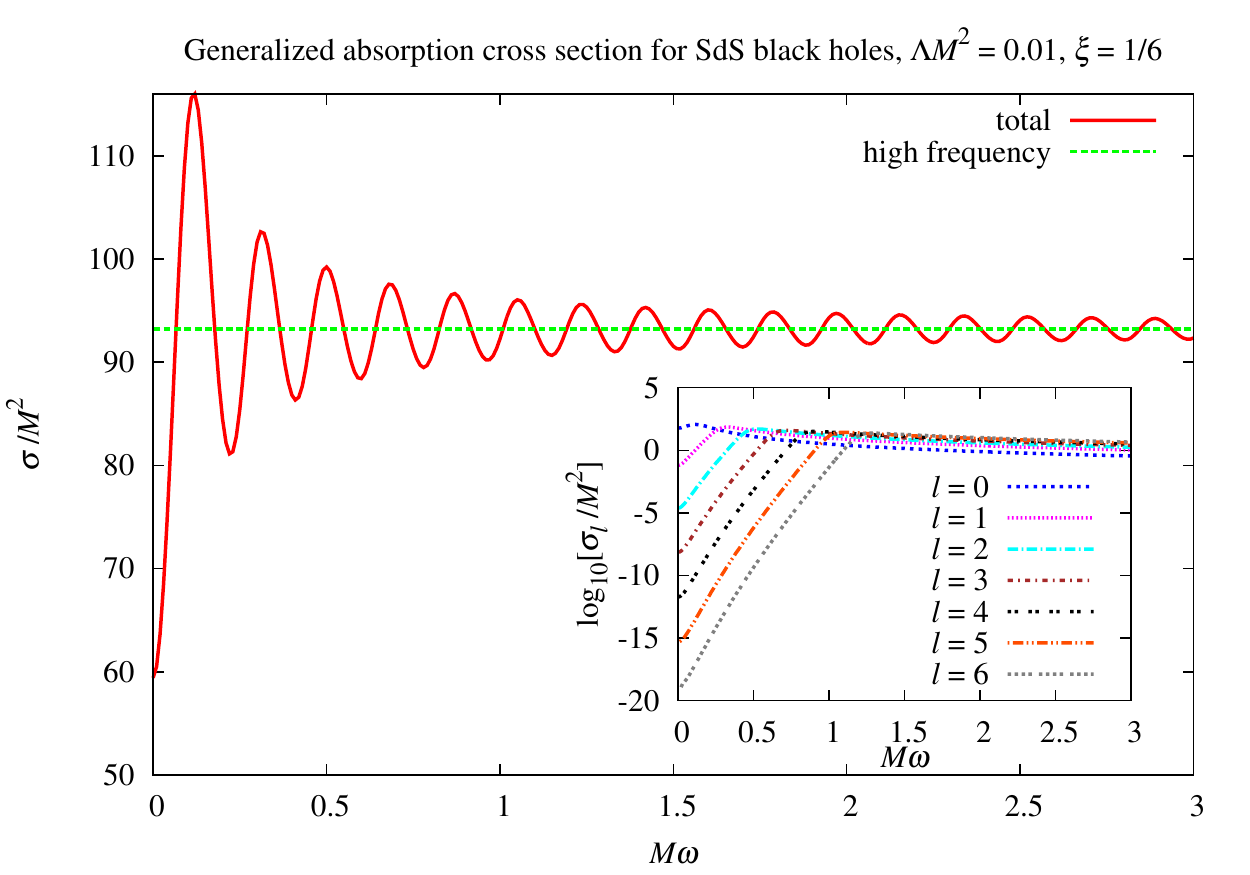}
 \caption{Total and partial generalized absorption cross sections
 for $\Lambda M^2 = 0.01$ and conformally coupled scalar field,
 $\xi = 1/6$, considering contributions of the angular momentum
 up to $ l = 20$.
 The total generalized absorption cross section is in
 excellent agreement with the high-frequency limit. The partial 
 contributions are shown in the log plot. 
 One can see that, although finite, the partial
 generalized absorption cross sections at $\omega=0$ are very small for $l >0$,
 and get smaller for larger values of $l$.}
 \label{acs_lem2x1b6}
\end{figure}

In FIG.~\ref{acs_lem2x1b6}, we present the total and partial
generalized absorption cross sections for $\Lambda M^2 = 0.01$
and $\xi  = 1/6$ (conformal coupling). 
The sum in Eq.~\eqref{tacs} is performed up to $ l = 20$. 
It can be seen that
the partial generalized absorption cross sections are nonzero for
$\omega = 0$, and that they get smaller for bigger values of $l$, presenting
a value  $\sim 10^{-19} M^2$ already for $l = 6$. This is in
agreement with Eq.~\eqref{simplified}.
As in FIG.~\ref{nacs}, the total generalized absorption cross section 
exhibited in FIG.~\ref{acs_lem2x1b6}
presents an excellent agreement with the high-frequency result
given by Eq.~\eqref{hfacs}.
We note that the $\Lambda$ dependence in $\sigma_{hf}$ is not significant for this agreement.

\section{Conclusion}
\label{conclusion}

In this paper we have analyzed the greybody factor of Schwarzschild-de Sitter black holes for non-minimally coupled scalar fields. In particular, it was shown that the greybody factor in the zero-frequency regime is nonzero only for the minimally coupled (and massless) scalar field.
For all other couplings to the scalar curvature, which may equivalently be regarded as mass terms, the greybody factors tend to zero like $\omega^2$, irrespective of the value of the coupling parameter $\xi$. 
In this sense the minimally coupled case is special.

A non-vanishing greybody factor in the low-frequency regime implies a nonzero Hawking emission in the same limit.
For $\xi \neq 0$, however, the emission rate in this limit is always zero. There is nevertheless an enhancement in the emission rate but it only occurs at finite, albeit small, frequencies.

We have obtained numerical results that are in good agreement with the analytical low-$\xi$ approximations derived in the low-frequency regime.
The numerical results also match the analytical small-$\lambda$ approximation in the low-frequency regime for the case of small Schwarzschild-de Sitter black holes.
We observed that the coupling to the scalar curvature only has a significant effect on the emission rate at low frequencies, $\omega \lsim M^{-1}$, and that the above-mentioned enhancement is more pronounced for larger black holes (larger $\lambda$) but becomes less significant as the coupling parameter $\xi$ grows.

Finally, we developed a sensible notion of a genera\-lized absorption cross section in asymptotically de Sitter spacetimes and investigated its properties.
In this res\-pect we found an excellent agreement between its high-frequency behavior and a geometric-optics description.

\acknowledgments

The authors thank Vitor Cardoso, Jorge Casti\~neiras, and George Matsas for helpful discussions.
The authors are grateful to Conselho Nacional de Desenvolvimento Cient\'\i fico e Tecnol\'ogico (CNPq)-Brazil,
to Coordena\c{c}\~ao de Aperfei\c{c}oamento de Pessoal de N\'\i vel Superior (CAPES)-Brazil,
Funda\c{c}\~ao para a Ci\^encia e Tecnologia (FCT)-Portugal,
and Marie Curie action NRHEP-295189- FP7-PEOPLE-2011-IRSES for partial financial support.
A.~H.\ and L.~C.\ also acknowledge partial support from the Abdus Salam International Centre for Theoretical Physics
through the Visiting Scholar/Consultant Programme and Associates Scheme, respectively.
J.~V.~R. is supported by Funda\c{c}\~ao para a Ci\^encia e Tecnologia (FCT)-Portugal
through contract no. SFRH/BPD/47332/2008.
A.~H. and J.~V.~R. thank Universidade Federal do Par\'a (UFPA) in Bel\'em,
and L.~C. thanks Instituto Superior T\'ecnico (IST) in Lisboa,
for kind hospitality during the completion of this work.

\begin{appendix}


\section{Square-well potential}
\label{swp}

In order to understand the change in the low-frequency behavior of the 
transmission coefficient at certain values of parameters in the potential, 
such as $\xi$ for the potential~(\ref{potential}),
the scattering problem in a negative square-well potential is instructive.
Thus, we consider the scattering problem
\begin{equation}
-\frac{d^2\psi}{dx^2} + V(x)\psi(x)= \omega^2\psi(x),
\end{equation}
where
\begin{equation}
V(x) = \begin{cases} -\omega_0^2\, , & {\rm if}\,\,\, 0 < x < a,\\
0\, , & {\rm otherwise}.
\end{cases}
\end{equation}
The transmission coefficient $\gamma(\omega)$ can readily be found as
\begin{equation}
\gamma(\omega) = \frac{4\omega^2\Omega^2}
{4\omega^2\Omega^2\cos^2\Omega a
+(\omega^2 + \Omega^2)^2\sin^2 \Omega a},
\end{equation}
where we have defined $\Omega = \sqrt{\omega^2 + \omega_0^2}$.

If $\omega_0 a \neq n\pi$ for any integer $n$, then for small $\omega$ we have
\begin{equation}
\gamma(\omega) \approx \frac{4\omega^2}{\omega_0^2\sin^2\omega_0 a}.  \label{Bgamma}
\end{equation}
We can readily see that the $\omega=0$ solution which equals $1$ for $x<0$, takes the value
$\psi(x)= \cos \omega_0 a - \omega_0(x-a) \sin \omega_0 a$ for $x>a$.  With $d\psi/dx = s=-\omega_0\sin\omega_0 a$  we can write Eq.~(\ref{Bgamma})
as $\gamma(\omega) \approx 4\omega^2/s^2$ [see Eq.~(\ref{btogamma})].
If $\omega_0 a = n\pi$ for some integer $n$, then the $\omega=0$ solution which equals $1$ for $x<0$ takes the constant value $(-1)^n$ for $x>a$. For these cases we indeed find
$\gamma(\omega)\to 1$, which is a constant, in the limit $\omega\to 0$.

\end{appendix}



\begin{thebibliography}{20}

\bibitem{Supernovae}
A. G. Riess et al. 
Astron. J. 116, 1009 (1998);
S. Perlmutter et al., 
Astrophys. J. 517, 565 (1999).

\bibitem{Guth:1980zm} A.~H.~Guth,
  Phys.\ Rev.\ D {\bf 23}, 347 (1981).

\bibitem{Strominger:2001pn} A.~Strominger,
  JHEP {\bf 0110}, 034 (2001).

\bibitem{Begelman:2003xt} M.~C.~Begelman,
  Science {\bf 300}, 1898 (2003).

\bibitem{MiniBHs}
  S.~Dimopoulos and G.~L.~Landsberg,
  Phys.\ Rev.\ Lett.\  {\bf 87}, 161602 (2001);
  S.~B.~Giddings and S.~D.~Thomas,
  Phys.\ Rev.\ D {\bf 65}, 056010 (2002).

\bibitem{Doran_etal} C. Doran, A. Lasenby, S. Dolan, and I. Hinder,
Phys. Rev. D \textbf{71}, 124020 (2005).

 \bibitem{Dolan_etal} S. Dolan, C. Doran, and A. Lasenby,
Phys. Rev. D \textbf{74}, 064005 (2006).

 \bibitem{cohm} L. C. B. Crispino, E. S. Oliveira, A. Higuchi, and G. E. A. Matsas,
Phys. Rev. D \textbf{75}, 104012 (2007).

 \bibitem{Dolan1} S. R. Dolan,
 Classical Quantum Gravity \textbf{25}, 235002 (2008).

\bibitem{cdo2} L. C. B. Crispino, S. R. Dolan, and E. S. Oliveira,
Phys. Rev. Lett. \textbf{102}, 231103 (2009).

 \bibitem{cdo1} L. C. B. Crispino, S. R. Dolan, and E. S. Oliveira,
 Phys. Rev. D {\textbf 79}, 064022 (2009).

 \bibitem{cho} L. B. Crispino, A. Higuchi, and E. S. Oliveira,
 Phys. Rev. D \textbf{80}, 104026 (2009).

\bibitem{Myung:2003cn} Y.~S.~Myung and H.~W.~Lee,
  Classical Quantum Gravity  {\textbf 20}, 3533 (2003).

\bibitem{HNS_atmp14_727} T.~Harmark, J.~Nat\'ario, and R.~Schiappa,
  Adv. Theor. Math. Phys. \textbf{14}, 727 (2010).

\bibitem{prl78_417} S.~R.~Das, G.~Gibbons and S.~D.~Mathur,
  Phys. Rev. Lett. \textbf{78}, 417, (1997).

\bibitem{Atsushi} A. Higuchi,
  Classical Quantum Gravity \textbf{18}, L139 (2001); \textbf{19}, 599 (2002).

\bibitem{prd13_198} D. N. Page,
  Phys. Rev. D \textbf{13}, 198 (1976).

\bibitem{Chen:2010ru} S.~Chen and J.~Jing,
  Phys.\ Lett.\ B {\bf 691}, 254 (2010).

\bibitem{BCKL} P.~R.~Brady, C.~M.~Chambers, W.~Krivan and P.~Laguna,
  Phys. Rev. D \textbf{55}, 7538 (1997).

\bibitem{KGB} P.~Kanti, J.~Grain and~A. Barrau,
  Phys. Rev. D \textbf{71}, 104002 (2005).

\bibitem{Liu:2010ar} M.~Liu, B.~Yu, R.~Wang and L.~Xu,
  Mod.\ Phys.\ Lett.\  {\bf 25}, 2431 (2010).

\bibitem{Wu:2008rb} S.~-F.~Wu, S.~-y.~Yin, G.~-H.~Yang and P.~-M.~Zhang,
  Phys.\ Rev.\ D {\bf 78}, 084010 (2008).

\bibitem{footnote} The `distance' between the two horizons is more naturally measured 
in the tortoise coordinate, in which the scattering problem in BH spacetimes takes the 
form of a one-dimensional Schr\"odinger problem. In this case the distance between
 the two horizons is infinite.

\bibitem{Gibbons:1977mu} G.~W.~Gibbons and S.~W.~Hawking,
  Phys.\ Rev.\ D {\bf 15}, 2738 (1977).

\bibitem{Stuchlik:1999qk} Z.~Stuchlik and S.~Hledik,
  Phys.\ Rev.\ D {\bf 60}, 044006 (1999).

\bibitem{Molina:2003ff} C.~Molina,
  Phys.\ Rev.\ D {\bf 68}, 064007 (2003).

\bibitem{BD} N. D. Birrell and P. C. W. Davies, Quantum Fields in Curved Space
(Cambridge University Press, Cambridge, England, 1982).

\bibitem{Starobinsky:1973} A.~A.~Starobinsky,
  Zh.\ Eksp.\ Teor.\ Fiz.\ {\bf 64}, 48 (1973) [Sov.\ Phys.\ JETP {\bf 37}, 28 (1973)];
      A.~A.~Starobinsky and S.~M.~Churilov,
  Zh.\ Eksp.\ Teor.\ Fiz.\ {\bf 65}, 3 (1973) [Sov.\ Phys.\ JETP {\bf 38}, 1 (1973)].

\bibitem{Unruh:1976} W.~G.~Unruh,
  Phys.\ Rev.\ D {\bf 14}, 3251 (1976).

\bibitem{Cardoso:2004hs} V.~Cardoso and O.~J.~C.~Dias,
  Phys.\ Rev.\ D {\bf 70}, 084011 (2004).

\bibitem{Candelas80} P.~Candelas, Phys.\ Rev.\ D {\bf 21}, 2185 (1980).

\bibitem{HiguchiCQG87} A.~Higuchi, Classical Quantum Gravity {\textbf 4}, 721 (1987).

\bibitem{Wald} R. M. Wald, \textit{General Relativity}
 (The University of Chicago Press, Chicago, 1984).

\bibitem{prd76_107502} L. C. B. Crispino, E. S. Oliveira, and G. E. A.
Matsas, Phys. Rev. D \textbf{76}, 107502 (2007).

\bibitem{Cohen} C. C.-Tannoudji, B. Diu, and F. Lalo\"e, 
\textit{Quantum Mechanics} vol. II (John Wiley \& Sons, New York, 1977).

\bibitem{Sanchez} N. S\'anchez,
  Phys. Rev. D \textbf{16}, 937 (1977).

\bibitem{Gradshteyn} I. S. Gradshteyn and I. M. Ryzhik, \textit{Table
of Integrals, Series, and Products} (Academic Press, San Diego,
2000), 6th ed.

\bibitem{footnote2}
For separable solutions of the form~\eqref{Phi}, equations~\eqref{KG} and~\eqref{radial_eq} are equivalent, but note that the first is obtained upon dividing the latter by a factor proportional to $f(r)$, which vanishes at the cosmological horizon.

\bibitem{Oliveira} E. S. Oliveira, \textit{Espalhamento e
absor\c c\~ao de campos bos\^onicos por buracos negros est\'aticos e
an\'alogos} (Ph.D. thesis, Universidade de S\~ao Paulo, S\~ao Paulo, 2009).


\end{thebibliography}
\end{document}